\documentclass[aip,graphicx,amsmath,amssymb,reprint]{revtex4-1}

\usepackage{graphicx}
\usepackage{dcolumn}
\usepackage{bm}

\usepackage[utf8]{inputenc}
\usepackage[T1]{fontenc}
\usepackage{mathptmx}

\begin{document}

\title{Dynamic Excitations of Chiral Magnetic Textures} 

\author{Martin Lonsky}
\email{lonsky@illinois.edu}
\author{Axel Hoffmann}
\affiliation{Materials Research Laboratory and Department of Materials Science and Engineering, University of Illinois at Urbana-Champaign, Urbana, Illinois 61801, USA}

\date{\today}

\begin{abstract}
Spin eigenexcitations of skyrmions and related chiral magnetic textures have attracted considerable interest over the recent years owing to their strong potential for applications in information processing and microwave devices. The emergence of novel material systems, such as synthetic ferri- and antiferromagnets, the continuing progress in micro- and nanofabrication techniques, and the development of more sophisticated characterization methods will undoubtedly provide a further boost to this young particular line of research. This Perspective summarizes the most significant advances during the past years and indicates future directions of both theoretical and experimental work.      
\end{abstract}


\maketitle 

\section{Introduction} \label{INTRO}
The high-frequency (hundreds of MHz to GHz) gyrotropic modes of magnetic vortices have been studied intensively during the past two decades.\citep{Guslienko2002, Park2003, Choe2004, Novosad2005, Buchanan2005, Kasai2006, Lee2008, Jung2011, Lee2011, Vogel2011, Haenze2014, Behncke2018} These fundamental excitations of the vortex ground state can, for example, be harnessed in spin-torque oscillators,\citep{Pribiag2007, Dussaux2010, Locatelli2011} spin-wave emitters \citep{Wintz2016} or magnonic crystals.\citep{Han2013, Shibata2004} Aside from the aforementioned resonant vortex dynamics, spin eigenexcitations of other chiral magnetic textures, particularly magnetic skyrmions,\citep{Roesler2006, Muhlbauer2009} may offer even more promising prospects towards the design of novel devices. This Perspective focuses on the most recent developments along this line since the appearance of the comprehensive review article by Garst \textit{et al.}\citep{Garst2017} It should also be noted that further, more exhaustive reviews on skyrmions include brief sections on resonant properties of these topological magnetic solitons.\citep{Mochizuki2015, Finocchio2016, Garst2016, Fert2017, Jiang2017, Everschor-Sitte2018, Back2020} The present article aims to provide an overview of the various excitation mechanisms of skyrmion resonance modes, to discuss the possible coupling of these eigenexcitations to different physical quantities or external perturbations, and to examine the major challenges and prospects with regard to the implementation of future experiments and the design of next-generation devices.
  
\begin{figure*}
\centering
\includegraphics[width=18 cm]{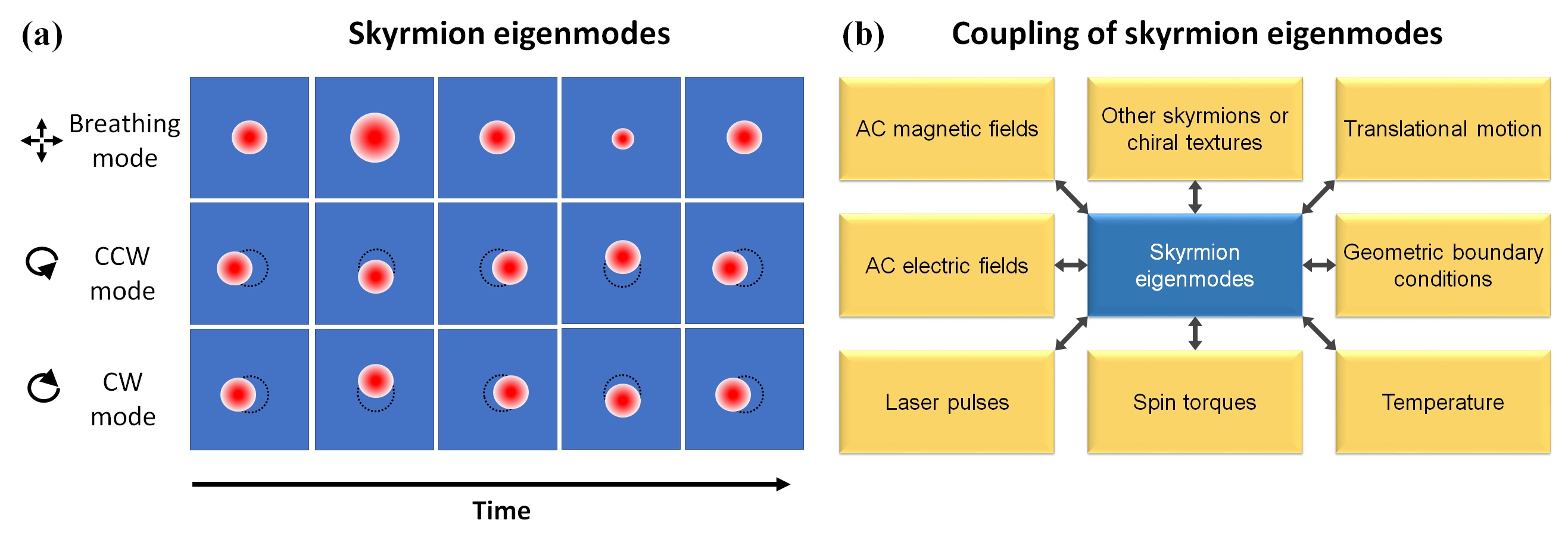}%
\caption{(a) Schematic temporal snapshots of three characteristic skyrmion eigenexcitations, namely the breathing mode as well as the CCW and CW gyration modes. Red and blue colors indicate opposite directions of out-of-plane magnetization. (b) Resonant skyrmion dynamics can be excited by various mechanisms and couple to different external perturbations such as time-varying fields or spin torques. Each scenario is discussed in detail throughout this Perspective.}
\label{MODECOUPLING}
\end{figure*}

Theoretical prediction of three characteristic skyrmion resonance modes that can be excited by microwave magnetic fields had been made by Mochizuki in the year 2012,\citep{Mochizuki2012} and was soon complemented by an experimental observation in the helimagnetic insulator Cu$_2$OSeO$_3$ using microwave absorption spectroscopy.\citep{Onose2012, Okamura2013} Later on, collective spin excitations of skyrmions were also observed in semiconducting and metallic helimagnets,\citep{Schwarze2015} and it was demonstrated that optical pump-probe techniques constitute an alternative experimental approach for the detection of these dynamic modes.\citep{Ogawa2015} 
The three types of magnetic resonances seen in experiments so far are termed breathing, clockwise (CW) and counterclockwise (CCW) modes. While the breathing mode entails a periodic oscillation of the skyrmion diameter, in the case of the other two resonances a CW or CCW gyrotropic motion is exercised by the skyrmion.\citep{Buettner2015} The latter two modes are similar to the characteristic eigenexcitation of a magnetic vortex, whereby the polarity of the vortex core determines the chirality of its gyrotropic motion. A schematic illustration of the three resonant modes of a single skyrmion is displayed in Fig.\ \ref{MODECOUPLING}(a). 
Apart from the three aforementioned resonance modes, theoretical studies predict the existence of further, higher magnetic multipole excitations of skyrmions which have not been observed in experiments up until now.\citep{Lin2014, Schuette2014} 
More generally, it should be emphasized that a major portion of published research work on dynamic excitations of chiral magnetic textures is of theoretical nature, while the number of reported experiments still remains low. The main reasons for this imbalance are comparably high magnetization dynamic damping parameters in materials utilized for magnetic films or multilayers that host skyrmions, and challenges related to the well-controlled generation and manipulation of skyrmions in micro- or nanoscale geometries. Furthermore, a large fraction of relevant thin-film or magnetic-multilayer samples are fabricated by means of sputter deposition, where often structural inhomogeneities, disorder and defects present additional complications. Consequently, this can also imply local variations of magnetic parameters such as the anisotropy and chiral exchange interactions. Even though these spatial variations are typically rather small, the high sensitivity of the skyrmions' static and dynamic properties on the interplay of different competing micromagnetic energy terms leads to difficulties with regard to the implementation of experiments involving dynamic excitations of these chiral magnetic textures. Therefore, a major challenge towards the realization of prospective experiments, or, even further, with regard to the design of reliable devices, will be the selection of appropriate materials, and the optimization of growth methods and parameters.
In addition, for the specific case of an individual skyrmion in a constricted geometry, a typical requirement is that the lateral size of the sample (for example, a magnetic nanodisk\citep{Kim2014, Mruczkiewicz2017, Mruczkiewicz2018}) needs to be comparable to the skyrmion diameter, which usually lies well below the micron scale and thus represents a further practical obstacle.      

As illustrated in Fig.\ \ref{MODECOUPLING}(b), there exist a variety of ways how skyrmion eigenmodes can be excited and how they can couple to different external perturbations. In the pioneering theoretical work by Mochizuki it was demonstrated that breathing and gyration modes of skyrmions can be excited by out-of plane and in-plane microwave magnetic fields, respectively.\citep{Mochizuki2012} As will be discussed in the course of this Perspective, other excitation mechanisms involve ac electric fields, laser pulses, spin torques, thermal energy, geometric boundary conditions, and the coupling to other adjacent skyrmions and their resonance modes. The remainder of this paper is structured as follows. In Sec.\ \ref{COUPLINGSKYRMIONS}, the coupling among skyrmions will be discussed for different geometries and materials. Section \ref{COUPLINGTRANSLATION} is devoted to the relationship between spin excitations and the translational motion of skyrmions. This is followed by a comprehensive overview of dynamic excitations of chiral magnetic textures other than skyrmions in Sec.\ \ref{OTHERTEXTURES}. Subsequently, the focus is on other experimental approaches such as the excitation of skyrmion resonance modes by means of spin torques or laser pulses, see Sec.\ \ref{OTHEREXPS}. Finally, the key points of the Perspective are summarized and we will provide a future outlook in Sec.\ \ref{SUMMARY}.  
  
\section{Coupling Among Skyrmions} \label{COUPLINGSKYRMIONS}
For several types of materials and geometries, it has been demonstrated that the coupling among adjacent skyrmions may allow for the design of novel information processing devices. Such devices could involve the energy-efficient transmission of information along straight and curved nanostrips in which no direct motion of skyrmions is required. Instead, the central mechanism for information transmission is given by the propagation of characteristic skyrmion eigenmodes.  

\begin{figure*}
\centering
\includegraphics[width=15.5 cm]{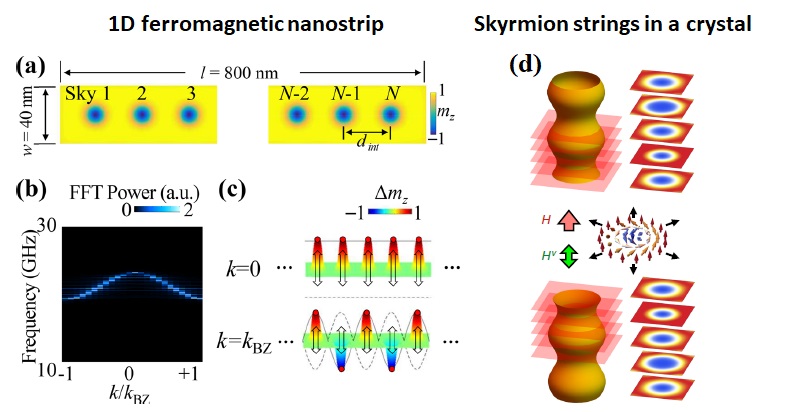}%
\caption{(a) One-dimensional skyrmion lattice in a thin ferromagnetic nanostrip. (b) Dispersion curve of the coupled skyrmion breathing modes. (c) Spatial profiles of $\Delta m_{z}$ components of individual skyrmions at $k=0$ and $k=\pi /a$ where $a=32\,$nm. Panels (a)-(c) have been reproduced with permission from Kim \textit{et al.}, J.\ Appl.\ Phys.\ \textbf{123}, 053903 (2018). Copyright 2018 AIP Publishing. (d) Schematic illustration of coupled breathing modes in a skyrmion string. Upper and lower parts correspond to snapshot images describing how the spin excitation can propagate along the skyrmion string. Reproduced from Seki \textit{et al.}, Nat.\ Comm.\ \textbf{11}, 256 (2020), under the terms of license CC BY 4.0.}
\label{COUPLEDMODES}
\end{figure*} 
Numerical simulations by Kim \textit{et al.}\ show that coupled breathing modes of skyrmions periodically arranged in a one-dimensional nanostrip exhibit a characteristic dispersion relationship which is controllable by an external perpendicular magnetic field that alters the inter-skyrmion distance.\citep{Kim2018} This scenario is depicted in Fig.\ \ref{COUPLEDMODES}(a)--(c). Figure \ref{COUPLEDMODES}(a) shows a schematic illustration of the considered skyrmion lattice with $N=25$ being the highest number of simulated skyrmions. The dispersion relation shown in Fig.\ \ref{COUPLEDMODES}(b) was determined after applying a magnetic field pulse only to the skyrmion at the left end of the nanostrip and calculating the fast Fourier transform of the temporal variation of the spatially-averaged $z$-component of the magnetization, $\langle m_{z} \rangle$, for each of the $25$ skyrmion core regions. The spatial profiles of $
\Delta m_{z}$ at two selected wave vectors, $k=0$ and $k=\pi /a$ (where $a=32\,$nm), are shown in Fig.\ \ref{COUPLEDMODES}(c). 
Aside from the previously described work, further numerical calculations by Zhang \textit{et al.}\ with regard to the spin-wave propagation along a linear skyrmion array lead to comparable results.\citep{Zhang2019} Furthermore, the authors also discuss the case of relatively large inter-skyrmion spacing, where the coupled breathing modes become less relevant and instead only the properties of the high-frequency spin waves related to the magnetization dynamics of the ferromagnetic nanostrip are important. More precisely, this high-frequency band is broken into discrete curves due to the periodically arranged skyrmions and thus works as a magnonic crystal. Consequently, these versatile dynamic properties of skyrmion lattices are promising for future magnon-spintronics devices.\citep{Wang2020, Ma2015, Ma2015a}       
    
Further comprehensive studies have been conducted for the case of coupled skyrmion gyration modes in a nanostrip \citep{Kim2017}, breathing and gyration modes in a chain of edge-to-edge connected skyrmion nanodots \citep{Mruczkiewicz2016}, as well as for propagating gyration modes of magnetic vortices in touching nanodisks.\citep{Han2013, Han2013a, Jeong2014} For the case of skyrmions in a linear chain of ferromagnetic nanodots, Gareeva and Guslienko have demonstrated that the characteristics of the collective excitation spectra depend on whether N\'{e}el or Bloch skyrmions are considered.\citep{Gareeva2018} 
A central parameter for the performance of potential devices based on coupled dynamic modes is the propagation speed of these excitations. The highest theoretically predicted propagation speed so far ($200$--$700\,$m/s) has been reported for coupled skyrmion breathing modes in a one-dimensional nanostrip.\citep{Kim2018} In comparison, the average speed for propagating skyrmion-gyration signals in such a nanostrip has been calculated to be $135\,$m/s,\citep{Kim2017} which is still more than twice as high as for vortex-gyration-signal propagation.\citep{Han2013a} However, in contrast to the propagating vortex excitations which have been studied experimentally by means of time-resolved scanning transmission x-ray microscopy,\citep{Han2013} no experimental work on propagating skyrmion eigenexcitations in one- or two-dimensional structures has been reported so far. As already mentioned in Sec.\ \ref{INTRO}, the lack of experiments can be explained by high damping parameters in relevant materials as well as difficulties in the well-controlled generation and arrangement of skyrmions in the considered geometries. In this context, further progress in materials engineering and a better control of skyrmions in thin-film micro- and nanostructures will be fundamental prerequisites towards the realization of experiments or even applications.   
    
Aside from such nano- or microstructured geometries based on ferromagnetic thin films, an alternative environment for propagating dynamic excitations are so-called skyrmion strings (also commonly referred to as skyrmion tubes) in three-dimensional systems.\citep{Lin2019, Seki2020, Xing2020, Birch2020} As an analog of vortex lines in superfluids \citep{Sonin1987} or type-II superconductors \citep{GENNES1964}, skyrmion strings are composed of uniformly stacked two-dimensional skyrmions along the string direction. These strings have been demonstrated to host characteristic propagating dynamic modes that are comparable to Kelvin modes in superfluids, where vibrations of a quantized vortex propagate along a vortex line.\citep{Lin2019, Sonin1987} A similar geometry has been studied by Ding \textit{et al.}\ for magnetic-vortex core strings in relatively thick Ni$_{80}$Fe$_{20}$ nanodots.\citep{Ding2014} However, in this case the detected resonances are not propagating along the vortex line, but instead correspond to standing modes along the dot thickness.     
In contrast to the previously discussed one- and two-dimensional systems, a comprehensive experimental study on the propagation dynamics of spin excitations along skyrmion strings has been carried out by Seki \textit{et al.}\ for the chiral magnetic insulator Cu$_{\mathrm{2}}$OSeO$_{\mathrm{3}}$.\citep{Seki2020} In this work, the coupled eigenexcitations were studied in the low-temperature skyrmion phase of a Cu$_{\mathrm{2}}$OSeO$_{\mathrm{3}}$ single crystal by means of propagating spin-wave spectroscopy. Owing to the comparably low Gilbert damping parameter ($0.005\leq \alpha \leq 0.05$ for different temperatures and various spin excitation modes) of this material, the breathing and gyration modes can propagate over a distance exceeding $50\,\mathrm{\mu m}$, which corresponds to a value more than $10^{3}$ larger than the diameter of a single skyrmion string. Moreover, the propagation speed can reach values up to $7000\,$m/s and thus is ten times higher than the predicted velocities for coupled breathing modes of skyrmions in one-dimensional nanostrips. On the other hand, disadvantages of the investigated material Cu$_{\mathrm{2}}$OSeO$_{\mathrm{3}}$ include the necessity of low temperatures and, in comparison to customizable thin-film geometries, a lower flexibility in the design of potential device architectures. An illustration of coupled breathing modes propagating along a skyrmion string is presented in Fig.\ \ref{COUPLEDMODES}(d). Here, it is assumed that the local spin eigenexcitations are launched at $z=0$. The local skyrmion oscillation at this position is indicated in the central part of panel (d). The upper image depicts the case of the wave vector $\textbf{k}$ pointing along the same direction as the external dc magnetic field $\textbf{H}$, while the lower section implies an antiparallel alignment of $\textbf{k}$ and $\textbf{H}$. The authors find that the propagation of these excitations is non-reciprocal and thus the dispersion is asymmetric.     

In addition to the previously described systems that host propagating skyrmion eigenmodes, the coupling among skyrmions (or their resonant dynamics) is also relevant in magnetic multilayers, such as synthetic ferri- and antiferromagnets. The latter structures consist of at least two ferromagnetic layers that are coupled antiferromagnetically via a nonmagnetic spacer layer. Only recently the stabilization of skyrmions in synthetic antiferromagnets has been demonstrated experimentally by Legrand \textit{et al.}\citep{Legrand2020} Comprehensive micromagnetic simulations have provided a first overview of the coupling between gyration\citep{Xing2018} or breathing\citep{Lonsky2020} modes in synthetic antiferromagnet trilayers. Moreover, numerical simulations by Dai \textit{et al.}\ have demonstrated a non-linear, flower-like dynamics for coupled skyrmion gyration modes in Co/Ru/Co nanodisks.\citep{Dai2014} 

The experimental detection of resonant skyrmion dynamics in magnetic multilayers is highly challenging, which is why here research is still in its infancy. Montoya \textit{et al.}\ performed temperature- and field-dependent ferromagnetic resonance measurements on dipole skyrmions in amorphous Fe/Gd multilayers and support their results by micromagnetic modeling.\citep{Montoya2017} The application of either in-plane or out-of-plane microwave magnetic fields leads to the excitation of four localized spin-wave modes similar to those of skyrmions that are stabilized by the Dzyaloshinskii-Moriya interaction. Furthermore, the authors demonstrate that the properties of the observed resonance spectra are not influenced by the precise arrangement of dipole skyrmions, that is, whether close-packed lattices or isolated skyrmions are considered. Lastly, it should be emphasized that the effective Gilbert damping parameter $\alpha_{\mathrm{eff}}$ has been found to lie below $0.02$, which is remarkably low for such a multilayer system.     
Another experimental study on resonant skyrmion dynamics in magnetic multilayers includes ferromagnetic resonance spectroscopy measurements of Ir/Fe/Co/Pt stacks.\citep{Panagopoulos2020} In detail, the gyrotropic modes of individual skyrmions in these multilayers with dipolar interlayer coupling could be detected. This is especially noteworthy in view of the fact that the determined effective damping parameter $\alpha_{\mathrm{eff}}\approx 0.1$ is more than five times higher compared to the previously discussed work by Montoya \textit{et al.}, which is mainly due to the summation of different effects such as spin wave scattering at interfaces and spin pumping.
Such high damping parameters make the detection of skyrmion eigenexcitations in magnetic multilayers considerably more difficult than for other systems such as single crystals. On the other hand, the fact that, despite the relatively strong damping, resonant skyrmion dynamics could be observed in Ir/Fe/Co/Pt multilayers is encouraging with regard to further experiments on other systems.        

In all of the above-mentioned cases, the resonant dynamics strongly depends on various adjustable parameters such as the interlayer exchange interaction or the external magnetic field. 
Therefore, the dynamic fingerprint of skyrmions in such magnetic multilayers could enable a detailed characterization of skyrmion states. For instance, measurements of the (anisotropic) magnetoresistance modulated by periodic breathing oscillations could offer a means for electric skyrmion detection in compensated multilayers, where due to the vanishing total magnetization other magnetic or electric sensing methods are highly challenging. For the case of magnetic vortices, the sensitive detection of resonance modes using homodyne voltage signals generated due to the anisotropic magnetoresistance has already been successfully demonstrated in experiments.\citep{Lendinez2020, Gangwar2015, Cui2015} Moreover, Penthorn \textit{et al.}\ have confirmed the presence of a single skyrmion in a magnetic tunnel junction by performing homodyne-detected spin-transfer driven magnetic resonance measurements.\citep{Penthorn2019} Characteristic signatures in the experimentally determined resonance spectrum were interpreted as the breathing mode of a skyrmion. The authors could reinforce this hypothesis by means of micromagnetic simulations. In conclusion, similar experiments are expected to be carried out in the future for other systems hosting magnetic skyrmions, such as thin-film micro- and nanostructures or magnetic multilayers.   

Finally, we note that aside from the aforementioned case of magnetic multilayers, clusters of skyrmions in ferromagnetic nanodisks are likewise expected to exhibit a characteristic dynamic fingerprint of correlated resonance modes that could be detected in experiments.\citep{Liu2020} 
Consequently, it is conceivable that, owing to the unique and complex signature of the resonant skyrmion dynamics, the application of broadband microwave impedance spectroscopy or related measurement techniques may constitute a promising all-electrical approach for the quantification of skyrmion densities in solids. A previous theoretical study dealing with the spectral analysis of topological defects in artificial spin-ice lattices clearly demonstrates the potential of such an approach.\citep{Gliga2013}   
 
\section{Interplay Between Spin Excitations and Translational Motion of Skyrmions} \label{COUPLINGTRANSLATION}
In this section, we will discuss various possibilities for the interplay between eigenmodes and the translational motion of skyrmions. While this type of coupling may be beneficial for certain applications, it will also be shown that there exist cases where the excited dynamic modes are considered a limiting factor for the efficient manipulation of skyrmions.  

In a recent theoretical work, Leliaert \textit{et al.}\ simulated the skyrmion motion through a nanotrack with periodically arranged constrictions, that is, areas with reduced Dzyaloshinskii-Moriya interaction.\citep{Leliaert2018} Such a geometry is potentially relevant for the realization of racetrack memory devices. If the skyrmion velocity and inter-notch distance are attuned to each other such that the time required for a skyrmion to move between two adjacent constrictions corresponds to a (small) multiple of the breathing mode period, the resonant oscillation of the skyrmion diameter is excited. In other words, the resonant velocity to excite a breathing mode can be controlled by varying the separation between the constrictions. The authors discuss that such a periodically notched nanotrack allows for stabilizing the skyrmion velocity against undesired perturbations, such as Oersted fields or random thermal motion. It is emphasized that notches with a weak potential are sufficient, since the breathing mode excitation might otherwise be so strong that the skyrmion collapses. Furthermore, it has also been demonstrated that the coupling between the skyrmion breathing mode and the velocity is present for both spin-transfer-torque and spin-orbit-torque driven motion. Up until now, such a scenario has not been realized experimentally.

Besides the previously discussed case of a nanotrack with periodically arranged constrictions, numerical calculations by Tomasello \textit{et al.}\ and Wang \textit{et al.}\ show that even for the current-induced skyrmion motion in a regular racetrack geometry transient breathing and gyration modes can be excited.\citep{Tomasello2014, Wang2019b} Both studies utilized a combination of micromagnetic modeling and an analytical description derived from the Thiele equation. Tomasello \textit{et al.}\ identify the excitation of breathing modes at high electric currents as a possible limitation of the maximum skyrmion velocity in racetrack memories. More specifically, due to the increasing magnitude of the diameter oscillations towards higher applied electric currents, the skyrmion state tends to break down at a certain threshold current density.\citep{Tomasello2014} In addition to that, Wang \textit{et al.}\ have reported that also gyration modes can be excited in such a scenario. In fact, these gyration modes are directly coupled to the breathing modes of the skyrmion which is manipulated by an electric current. In conclusion, resonant skyrmion dynamics can also be (unintentionally) excited in the scenario of current-induced skyrmion manipulation in racetrack devices and thus represents a limiting factor with regard to the maximum achievable velocity.      

By contrast, a further direction where the interplay between skyrmion eigenmodes and the translational motion offers great potential for applications is given by the microwave-driven skyrmion propagation under inclined static magnetic fields which break the rotational symmetry of the chiral magnetic texture. In detail, both electric and magnetic microwave fields have been shown to facilitate the efficient manipulation of skyrmions. 
As reported by Yuan \textit{et al.}, the combination of a perpendicularly oscillating electric field that modifies the local magnetic anisotropy (so-called parametric pumping) and thus can excite skyrmion breathing modes, and an in-plane dc magnetic field which breaks the rotational symmetry can cause a wiggling motion along a well-defined trajectory.\citep{Yuan2019} Comparable numerical studies by Song \textit{et al.}\ have found a trochoidal motion of antiskyrmions that is likewise induced by the combination of an in-plane magnetic field and perpendicular microwave electric fields.\citep{Song2019a} 
Further theoretical investigations have focused on skyrmion motion driven by ac magnetic fields in combination with inclined static magnetic fields.\citep{Ikka2018, Wang2015} Wang \textit{et al.}\ demonstrated for the first time that skyrmion propagation can be induced by breathing-mode excitation under magnetic fields that are inclined from the perpendicular direction to the skyrmion plane.\citep{Wang2015} The authors could show that both an individual skyrmion and a skyrmion lattice can be manipulated in this manner. 
Besides the well-established breathing and gyration modes, novel resonance modes which correspond to combinations of the former two were demonstrated by Ikka \textit{et al.}\citep{Ikka2018} In detail, the authors have shown for a two-dimensional skyrmion lattice that the direction and velocity of the skyrmion propagation strongly depend on the excited mode. For instance, the velocity is significantly higher for a mode that includes a strong counterclockwise-gyration component.

The highest predicted skyrmion propagation velocities in the aforementioned studies are as high as a few meters per second \citep{Yuan2019} and thus exceed those of temperature-gradient-driven motion.\citep{Kong2013} However, compared to the highest achieved velocities ($\sim 100\,$m/s) of skyrmions driven by short current pulses,\citep{Woo2016} the microwave-driven skyrmion propagation is comparably slow. It can be expected that optimizations of various materials parameters will lead to enhanced velocities. Note that, for example, the propagation speed exhibits a nontrivial dependence on the damping parameter $\alpha$.\citep{Yuan2019, Song2019a} Moreover, the experimental realization of the microwave-induced skyrmion propagation under tilted magnetic fields still needs to be demonstrated.              

Lastly, we point out a further notable effect that is predicted to occur in the previously described scenarios---however, it is not connected to the translational motion of skyrmions. Koide \textit{et al.}\ have theoretically shown for a skyrmion crystal that the microwave excitation of skyrmion eigenmodes in the presence of a tilted static magnetic field can also lead to the generation of a temporally oscillating spin-driven electromotive force or voltage with a large dc component.\citep{Koide2019} This spinmotive force can be regarded as the inverse effect of the spin-transfer torque mechanism. Consequently, the efficient conversion of microwaves to a dc electric voltage via skyrmion-hosting materials has been proposed for future spintronics devices.   

\section{Spin Eigenexcitations of Other Chiral Magnetic Textures} \label{OTHERTEXTURES}
In this section, we will discuss studies dealing with the spin eigenexcitations of other chiral magnetic textures that are strongly related to skyrmions. In particular, we are focusing on antiskyrmions, $k\pi$ skyrmions, irregularly shaped skyrmions and other objects.

\begin{figure*}
\centering
\includegraphics[width=18.0 cm]{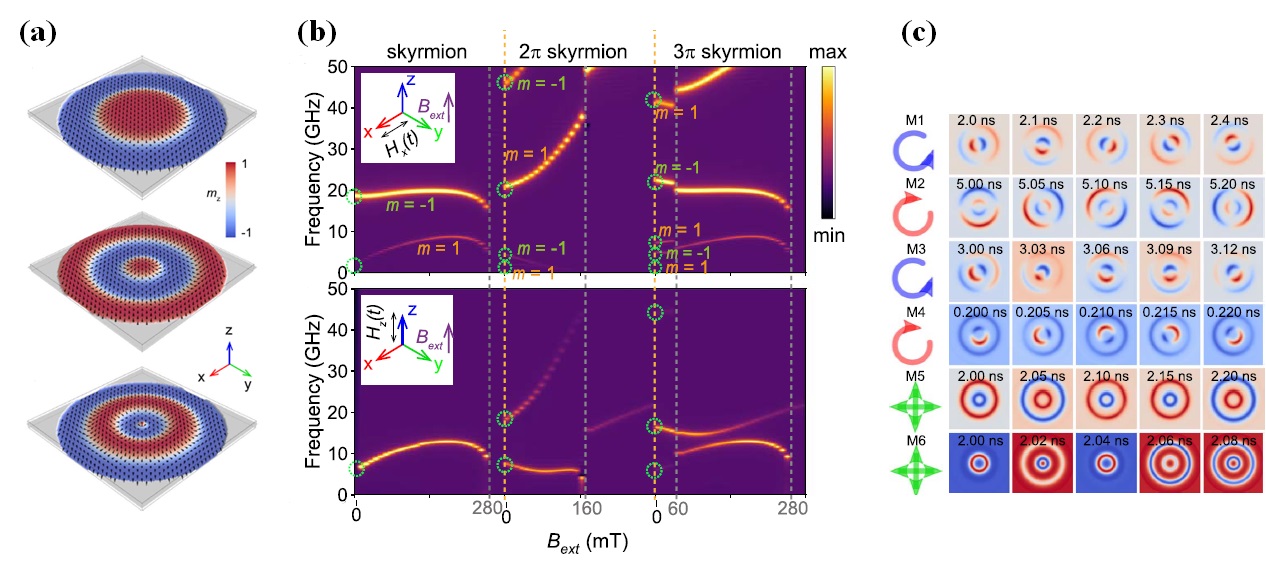}%
\caption{(a) Illustration of a regular $1\pi$ skyrmion, a $2\pi$ skyrmion and a $3\pi$ skyrmion in a nanodisk. (b) Contour plots of calculated power spectra for $k\pi$ skyrmions excited by an in-plane (top panel) and out-of-plane (bottom panel) ac magnetic field. Static magnetic fields in the range of $0$--$300\,$mT were applied. Gray dashed lines represent critical static field at which the magnetization state transforms to another one. (c) Snapshots of six excitation modes for a $2\pi$ skyrmion. M1 to M4 are gyration modes excited by an in-plane ac magnetic field, whereas M5 and M6 correspond to breathing modes induced by an out-of-plane time-varying field. Reproduced from Song \textit{et al.}, New.\ J.\ Phys.\ \textbf{21}, 083006 (2019), under the terms of license CC BY 3.0.} 
\label{KPISKYRM}
\end{figure*}   
Firstly, we will discuss two theoretical articles that have reported on excitation modes of so-called $k\pi$ skyrmions.\citep{Rozsa2018, Song2019} In addition to regular ($1\pi$) skyrmions, where the magnetization direction rotates by 180 degrees along the radial direction, a stabilization of $k\pi$ skyrmions can be achieved in constricted geometries.\citep{Bogdanov1999} For these objects, the magnetization rotates multiple times between the core and the collinear region. Here, $k$ indicates the number of sign changes of the out-of-plane magnetization component along the radial direction. A schematic illustration of the three lowest-order $k\pi$ skyrmions is displayed in Fig.\ \ref{KPISKYRM}(a). 
Note that these chiral textures are also referred to as target skyrmions. Moreover, the specific case of $2\pi$ skyrmions is termed a skyrmionium.\citep{Kolesnikov2018} While the existence of $k\pi$ skyrmions has been reported by several research groups during the past years,\citep{Zhang2018, Zheng2017, Cortes-Ortuno2019, Kent2019} their dynamic excitations still remain to be investigated experimentally. The theoretical studies which will be presented in this section suggest that the realization of corresponding experiments can be expected in the near future.

By using an atomistic classical spin model, R\'{o}zsa \textit{et al.}\ study the localized spin waves in isolated $k\pi$ skyrmions on a field-polarized background.\citep{Rozsa2018} A main result of this work is that the number of excitation modes increases towards higher values of $k$. This is explained by the additional excitations with higher angular momentum quantum numbers $m$ along the larger perimeter of the magnetic texture. Moreover, the authors have discussed possible instabilities of $k\pi$ skyrmions: While $2\pi$ and $3\pi$ skyrmions at low fields are destroyed through a burst instability, an elliptic instability is found for regular $1\pi$ skyrmions. By contrast, at high magnetic fields the innermost ring of the structure collapses for all $k\pi$ skyrmions due to an instability of the breathing mode.    
Similar results on the dynamic eigenexcitations of $k\pi$ skyrmions have been obtained by Song \textit{et al.}\ who performed extensive micromagnetic simulations.\citep{Song2019} In addition to the increasing number of resonances towards higher $k$ values, the authors have also studied the dependence of gyration- and breathing-mode eigenfrequencies on the external static magnetic field which implies an expansion or reduction of different parts of the $k\pi$ skyrmion. Selected results are presented in Fig.\ \ref{KPISKYRM}(b) and (c). Figure \ref{KPISKYRM}(b) shows contour plots of the simulated power spectra for the three lowest-order $k\pi$ skyrmions in a nanodot. The spectra were obtained by a fast Fourier transform of the corresponding magnetization components for static external magnetic fields between $0$ and $300\,$mT. The excitation of the $k\pi$ skyrmions has been simulated by applying in-plane (top panel) and out-of-plane (bottom panel) ac magnetic fields. Gray dashed lines represent the critical dc magnetic fields where the magnetization state collapses to another one. Clearly, these power spectra confirm the increasing number of resonance modes for higher-order $k\pi$ skyrmions. Snapshots of the six distinct modes for the case of a $2\pi$ skyrmion are depicted in Fig.\ \ref{KPISKYRM}(c) at five different points in time. While M1 to M4 represent gyration modes that can be excited by in-plane microwave magnetic fields, modes M5 and M6 correspond to two breathing modes that are driven by an out-of-plane field.      

In addition to the studies on $k\pi$ skyrmions, Booth \textit{et al.}\ focus on the more specific case of a skyrmionium ($2\pi$ skyrmion) in a nanodisk.\citep{Booth2019} It is pointed out that the three-dimensional character of the skyrmionium, that is, its finite thickness in the simulated nanodisk geometry, can imply a complex behavior in response to time-varying external magnetic fields. In the near future, other three-dimensional chiral magnetic textures such as skyrmion tubes (see also Sec.\ \ref{COUPLINGSKYRMIONS}), chiral bobbers \citep{Zheng2018} or magnetic hopfions \citep{Liu2020a, Wang2019a} are expected to be investigated in greater detail---including the characterization of their corresponding dynamic excitation modes.   

Aside from the fundamental importance of a systematic analysis of skyrmionium resonance modes, these eigenexcitations can also be relevant for the creation\citep{Vigo-Cotrina2020} or switching\citep{Vigo-Cotrina2020a} of a skyrmionium by means of time-varying magnetic fields. In detail, numerical calculations by Vigo-Cotrina and Guimar$\tilde{\mathrm{a}}$es show that skyrmioniums and skyrmions in Co/Pt nanodisks can be created by applying pulses of oscillating perpendicular magnetic fields with a frequency corresponding to the respective eigenfrequencies.\citep{Vigo-Cotrina2020} Interestingly, the results imply that the strength of the magnetic field as well as the pulse duration determine whether a skyrmionium or a skyrmion is created. As a consequence, the selective creation of these magnetic textures could be possible in applications. Similarly, oscillating perpendicular magnetic field pulses with a frequency equal to that of the spin wave mode can switch the polarity of such chiral magnetic objects.\citep{Vigo-Cotrina2020a}

Up until now, only skyrmions with a radial symmetry have been discussed in this article. However, it should be emphasized that breathing and gyration modes can also occur for irregularly shaped skyrmions.\citep{Rodrigues2018} Rodrigues \textit{et al.}\ have developed a theoretical model in which such deformed skyrmions are described as closed domain walls. This extended soft-mode formalism for domain walls provides an analytical description of the dynamic excitations of both deformed and circular skyrmions. Moreover, the authors note that the developed formalism is further extendable to more generalized types of Dzyaloshinskii-Moriya interaction and thus would cover the dynamics of other chiral magnetic objects like antiskyrmions. Compared to regular skyrmions, the winding number of antiskyrmions is opposite in sign. A recent theoretical work by McKeever \textit{et al.}\ indicates that the breathing dynamics of antiskyrmions can be described in analogy to regular skyrmions.\citep{McKeever2019} This has also been confirmed by Liu \textit{et al.}\ in micromagnetic simulations of antiskyrmions in ultrathin ferromagnetic nanodisks.\citep{Liu2020b} Furthermore, the latter work also demonstrates the existence of gyration modes for antiskyrmions. Minor differences in the resonance frequencies of antiskyrmion and skyrmion eigenexcitations are attributed to the disparate internal magnetic configurations of the two magnetic textures.        

In addition, Song \textit{et al.}\ have shown that---similar to the case of skyrmions discussed in Sec.\ \ref{COUPLINGTRANSLATION}---a trochoidal motion of antiskyrmions can be induced by microwave electric fields along with a symmetry-breaking in-plane magnetic field.\citep{Song2019a} As for skyrmions, the time-varying electric field gives rise to parametric pumping of magnetic anisotropy and thereby excites the breathing mode of the antiskyrmion. An illustration of such a scenario is displayed in Fig.\ \ref{ACELFIELDS}(a) for the case of a skyrmion. 

Another recent theoretical study by D\'{i}az \textit{et al.}\ considers the coupling between the gyrotropic CW and CCW modes in a skyrmion-antiskyrmion bilayer.\citep{Diaz2020a} In detail, both gyrating cores emit spin waves along preferred directions with dipole signatures in their radiation pattern. The exact patterns strongly depend on the topological charge and the helicity of the two chiral magnetic textures. The authors discuss that such a system could be relevant for the design of novel efficient spin-wave emitters with tunable characteristics.       

Lastly, the spin eigenexcitations of skyrmions in antiferromagnets will be discussed. For the case of a thin-film collinear uniaxial antiferromagnet, Kravchuk \textit{et al.}\ could demonstrate by means of analytical and numerical calculations that there exist two branches of resonance modes.\citep{Kravchuk2019} In detail, the low-frequency branch corresponds to the localized modes of a skyrmion in a ferromagnet, whereas modes of the high-frequency branch do not have direct ferromagnetic counterparts. The latter are situated in the vicinity of the magnon continuum and are not involved in the instability process of skyrmions. The low-frequency branch of skyrmions with a large radius can be described by a formalism in which the antiferromagnetic skyrmion is interpreted as circular antiferromagnetic domain walls. In other words, this is a direct extension of the above-mentioned method developed by McKeever \textit{et al.}\ for ferromagnetic skyrmions and antiskyrmions.\citep{McKeever2019}        
Experimental detection of the antiferromagnetic skyrmion eigenexcitations could be realized for example by utilizing Brillouin light scattering techniques. 

A further class of material systems that host antiferromagnetic skyrmions are synthetic antiferromagnets. These systems have already been discussed in Sec.\ \ref{COUPLINGSKYRMIONS}. Here, we emphasize that the dynamic behavior in such multilayer stacks is different than in crystalline antiferromagnets, since the interlayer exchange coupling is much weaker than the direct exchange. The stronger separation of the two magnetic subsystems in synthetic antiferromagnets allows for the presence of small dipolar fields.\citep{Lonsky2020} Consequently, the experimental characterization of skyrmion eigenexcitations in synthetic antiferromagnets  is expected to be less challenging than for the case of crystalline antiferromagnets. 
Finally, for synthetic ferrimagnets with a sufficiently high degree of imbalanced magnetic moments, micromagnetic simulations have indicated the existence of non-radial skyrmion eigenexcitations in addition to the regular breathing modes.\citep{Lonsky2020} Similar non-radial dynamic resonances have also been observed in numerical simulations of skyrmions in FePt multilayer films.\citep{Bi2020} In conclusion, further theoretical and experimental studies focusing on non-radial skyrmion excitation modes in magnetic multilayers can be anticipated in the near future.  
  
\section{New Experimental Approaches and Possibilities} \label{OTHEREXPS}
In this section, we will discuss several other experimental approaches and possibilities with regard to the excitation and exploitation of skyrmion eigenmodes. 

\subsection{Electric Fields and Spin Torques}
\begin{figure}
\centering
\includegraphics[width=8.5 cm]{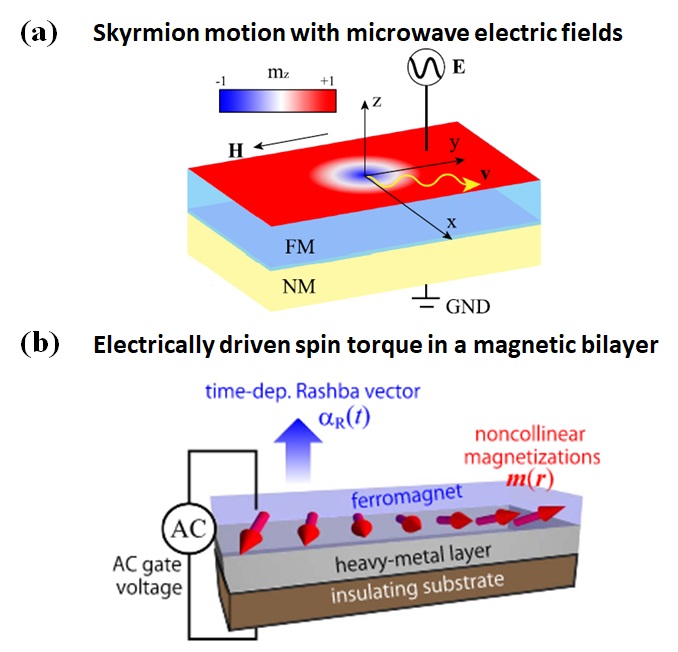}%
\caption{(a) Schematic illustration of a bilayer system composed of a ferromagnet (FM) and nonmagnetic (NM) layer. The combination of a microwave electric field along the $z$-direction and a symmetry-breaking in-plane magnetic field can induce a wiggling motion (yellow arrow) of the simulated skyrmion. Reproduced with permission from Yuan \textit{et al.}, Phys.\ Rev.\ B \textbf{99}, 014428 (2019). Copyright 2019 American Physical Society. (b) Time-dependent Rashba electron system interacting with local magnetization $\textbf{m(r)}$. The Rashba parameter $\alpha_{\mathrm{R}}(t)$ is time-modulated by an external ac gate voltage. Reproduced from Takeuchi \textit{et al.}, Sci.\ Rep.\ \textbf{9}, 9528 (2019), under the terms of license CC BY 4.0.} 
\label{ACELFIELDS}
\end{figure}   
As has been mentioned in Secs.\ \ref{COUPLINGSKYRMIONS} and \ref{OTHERTEXTURES}, the application of microwave electric fields allows for the excitation of skyrmion and antiskyrmion eigenmodes. In combination with a symmetry-breaking dc magnetic field this can drive the motion of the aforementioned chiral magnetic textures.\citep{Song2019a, Yuan2019} An exemplary geometry for a skyrmion in a bilayer is depicted in Fig.\ \ref{ACELFIELDS}(a).
The selective activation of isolated magnetic skyrmions in ferromagnets with microwave electric fields has also been discussed by Takeuchi and Mochizuki.\citep{Takeuchi2018} In detail, this theoretical work assumes an oscillatory modulation of the Dzyaloshinskii-Moriya interaction across the interface of a heterostructure by applying microwave electric fields, which eventually leads to the excitation of skyrmion eigenmodes. The authors point out that an advantage of ac electric fields over time-varying magnetic fields is that no gigantic ferromagnetic resonances are excited in the studied magnetic bilayer. Such strong ferromagnetic resonances would typically drown out the weaker skyrmion eigenmodes. Furthermore, the enhanced absorption of microwave energy would result in a low efficiency and considerable temperature rise. Similar findings have been reported in another theoretical work by Chen and Li.\citep{Chen2019} The experimental realization of the aforementioned scenario is still an open challenge. While, for instance, the voltage control of the Dzyaloshinskii-Moriya interaction has been successfully demonstrated for the case of dc electric fields in Ta/FeCoB/TaO$_{\mathrm{x}}$ trilayers,\citep{Srivastava2018} it remains unclear whether this would also be possible for microwave fields, since the underlying physical mechanism may involve a modulation of the oxygen stoichiometry at the interface. Thus it has yet to be investigated if the oxygen ion migration can take place on the desired fast time scales.

Apart from conventional microwave electric fields, Takeuchi \textit{et al.}\ have later theoretically studied the activation of skyrmion breathing dynamics by means of spin torques in a time-varying Rashba electron system driven by an ac gate voltage.\citep{Takeuchi2019} Such a resonant excitation of magnetic textures is possible since the considered ac torques act as an interfacial, temporally oscillating Dzyaloshinskii-Moriya interaction. Both isolated skyrmions and skyrmion lattices can be excited resonantly by this method. Figure \ref{ACELFIELDS}(b) shows a schematic illustration of the considered model system in which a torque arises from the time-varying Rashba-type spin-orbit interaction. As pointed out by Takeuchi \textit{et al.}, current issues towards the realization of devices are given by the enhanced complexity of spin-orbit interaction for candidates like ferromagnet/transition-metal-dichalcogenide or ferromagnet/topological-insulator systems, and which of the several relevant terms at the interfaces of such systems can be modulated by an ac gate voltage in reality.\citep{Takeuchi2019}

A simpler example for the resonant excitation of dynamic skyrmion modes by means of spin-transfer torques is discussed in a work by Dai \textit{et al.}, where the authors present micromagnetic simulations of coupled skyrmions in Co/Ru/Co nanodisks.\citep{Dai2016} More specifically, it is demonstrated that gyration modes can be selectively activated by applying spin-polarized ac currents with adequate frequency. In analogy to this observation, coupled skyrmion breathing modes in synthetic antiferromagnets can likewise be excited by spin-transfer torques.\citep{Lonsky2020}        

More generally, both spin-transfer torques and spin-orbit torques constitute promising tools to excite skyrmion dynamics in experiments or applications, such as skyrmion-based spin transfer nano-oscillators (STNOs).\citep{Sisodia2019a, Zhang2015, Garcia-Sanchez2016, Guslienko2020, Zhou2015} In contrast to conventional STNOs, lower threshold currents, high output power and narrow spectra linewidths can be expected for STNOs which involve the dynamic excitations of skyrmions.\citep{Sisodia2019a} Currently, a major portion of the proposed skyrmion-based STNOs is based on magnetic tunnel junctions with spin-transfer torques originating from spatially uniform, spin-polarized currents. 

The excitation of skyrmion eigenmodes by means of spin-orbit torques has not been investigated in great detail yet. So far, only Woo \textit{et al.}\ have reported on breathing-like dynamics excited by such torques.\citep{Woo2017} However, the authors point out that the observed behavior does not correspond to the internal skyrmion breathing modes. 
It should be emphasized that for the excitation of resonant skyrmion dynamics in magnetic films and multilayers with perpendicular anisotropy, novel types of spin-orbit torques that are generated in materials with reduced symmetry---for example, due to magnetic order or because of low crystalline symmetry---can be expected to constitute a new experimental approach.\citep{MacNeill2016, Baek2018, Safranski2018, Holanda2020, Liu2019} 
In conclusion, both theoretical and experimental works along this line are expected to be conducted in the near future.  

\subsection{Optical Excitation of Eigenmodes}
Thus far, the optical excitation of skyrmion eigenmodes has been reported solely in two articles. In a pioneering work, Ogawa \textit{et al.}\ have coherently excited the skyrmion lattice in Cu$_2$OSeO$_3$ by exploiting the impulsive magnetic field from the inverse-Faraday effect.\citep{Ogawa2015} In detail, the authors could identify both the breathing and the (CW and CCW) gyration modes of skyrmions in all-optical spin wave spectroscopy experiments. 

By contrast, Padmanabhan \textit{et al.}\ have successfully demonstrated the collective spin excitations of skyrmions in the lacunar spinel GaV$_4$S$_8$.\citep{Padmanabhan2019} Here, a different mechanism is responsible for the optical excitation, namely the laser-induced thermal modulation of the uniaxial anisotropy of the material---notice that the excitation mechanism in the previously mentioned study by Ogawa \textit{et al.}\ is nonthermal. In other words, the coupling between the laser pulse and the magnetic moments is mediated by the crystal lattice. Both the breathing mode and the CCW gyration mode have been detected by means of ultrafast time-resolved magneto-optical Kerr effect measurements.   

In the near future, the optical modulation of other magnetic interactions and the optical excitation of skyrmion eigenmodes in other materials is expected to provide further valuable insights into the GHz-range skyrmion dynamics. Furthermore, the optical excitation of related chiral magnetic textures is conceivable. 

\subsection{Thermal Effects and Mechanical Strain}
As has been demonstrated throughout this Perspective, there exist a plethora of micromagnetic modeling studies of static and dynamic properties of skyrmions. A majority of these simulations do not take into account any thermal effects, but instead assume absolute zero temperature. Even though this is an appropriate simplification for many cases, a finite temperature may play an important role in certain considerations, such as the lifetime and thermal stability of skyrmions. The precise calculation of lifetimes in a sufficiently wide temperature range is important with regard to the stability of individual skyrmion bits in devices.   
In this context, Desplat \textit{et al.}\ have found that the respective internal dynamic modes are highly relevant for the thermally-activated annihilation process of metastable skyrmions and antiskyrmions.\citep{Desplat2018} While there exist various paths to annihilation, the authors have shown that the most probable mechanisms are the skyrmion's collapse on a defect or its escape through a boundary. In both cases, different types of radially symmetric and distorted breathing modes are involved in the annihilation mechanism. Similar results have been reported by Sisodia and Muduli for the thermal decay of a single N\'{e}el skyrmion in a nanodisk.\citep{Sisodia2019} In detail, the presence of stochastic thermal noise leads to the excitation of gyration and breathing modes. Interestingly, the skyrmion changes its helicity with the same frequency as that of the excited breathing mode. Under certain conditions (higher temperatures and large skyrmion radius), a helicity slip is observed. Furthermore, the skyrmion decays at higher temperature via a mixed N\'{e}el-vortex state.

Apart from thermal effects it has been demonstrated that mechanical strain may also lead to the creation, deformation, transformation or annihilation of magnetic skyrmions. For instance, calculations by Hu \textit{et al.}\ suggest that a skyrmion state can switch to a variety of other magnetic configurations under biaxial in-plane strains.\citep{Hu2020} Such strains can be induced by applying an electric voltage across a piezoelectric layer located beneath the skyrmion-hosting magnetic film.  
The resulting magnetic states can be a single-domain or a vortex configuration under isotropic strains, and an elliptical skyrmion or a stripe domain under anisotropic strains. Interestingly, the strain-induced transition from a skyrmion to a single-domain state is accompanied by the excitation of a breathing mode. Moreover, the authors discuss that, since biaxial isotropic in-plane strains are equivalent to an out-of plane effective magnetic field, in principle the application of a dc magnetic field may be sufficient to excite resonant breathing dynamics. However, as for the strain-induced dynamics, this has not yet been demonstrated experimentally.   

\subsection{Alternative Techniques for the Detection of Internal Skyrmion Modes}
The limited number of works reporting on the experimental detection of skyrmion eigenmodes have mostly utilized the well-established microwave absorption spectroscopy or the above-discussed all-optical spin wave spectroscopy measurements\citep{Ogawa2015}. An alternative technique for the detection of resonant skyrmion dynamics has been established by P\"{o}llath \textit{et al.}\ in a recent article.\citep{Poellath2019} The technique is termed resonant elastic x-ray scattering ferromagnetic resonance (REXS-FMR) and combines the microwave excitation of dynamic modes by means of a coplanar waveguide with the detection of the scattered intensity in REXS measurements. As a proof of principle, the authors successfully identified the skyrmion resonance modes as well as other excitations in the field-polarized and helimagnetic phases of the chiral magnet Cu$_2$OSeO$_3$. 

Another experimental technique that has been developed recently relies on the detection of magnetic noise locally produced by thermal populations of magnons.\citep{Finco2020} This is achieved by performing relaxometry measurements with a quantum sensor based on a single nitrogen-vacancy defect in diamond. Even though in the case of the skyrmions considered by Finco \textit{et al.}\ the internal breathing mode frequencies are larger than the electron spin resonance frequency of the nitrogen-vacancy defect and thus the increased noise magnitude above the skyrmions is attributed to the scattering of propagating spin waves at these chiral textures, for other selected systems this technique may be sensitive to the resonant skyrmion dynamics. More generally, it should be emphasized that this type of quantum spin sensing allows to image non-collinear antiferromagnetic textures that are typically not detectable by other means.    

Aside from the aforementioned techniques, Brillouin light scattering (BLS) constitutes a further natural approach towards the characterization of spin-wave eigenmodes in skyrmion lattices or at individual skyrmions.\citep{Sebastian2015} Even though such experiments have not yet been reported, there are no fundamental reasons that BLS studies of resonant skyrmion dynamics would not be possible. A major advantage of BLS measurements could be the high spatial resolution of this technique.          

\section{Summary and Outlook} \label{SUMMARY}
In this Perspective, we have reviewed the experimental and theoretical efforts dealing with dynamic eigenexcitations of magnetic skyrmions and other chiral magnetic textures. Up to now the majority of published studies are of theoretical nature, whereas experiments are complicated by enhanced magnetization dynamic damping parameters in relevant materials as well as inhomogeneities in samples fabricated by sputtering. Therefore, intensified materials engineering efforts will be required in the near future in order to improve the quality of samples and devices. Aside from these apparent experimental difficulties, it has been demonstrated that the investigation and exploitation of skyrmion eigenexcitations has great potential for both fundamental studies and applications. Owing to the multiple excitation and coupling mechanisms of these resonance modes, it can be expected that this research field will experience a significant boost in the next years. One may envision information processing devices based on the propagation of skyrmion breathing or gyration modes, and, for instance, the practial realization of an all-electrical characterization of skyrmion states as suggested in Sec.\ \ref{COUPLINGSKYRMIONS}. Moreover, the excitation and detection of resonant skyrmion dynamics by additional techniques such as Brillouin light scattering will complement already existing experimental capabilities. Lastly, the experimental investigation of the resonant dynamics of other chiral magnetic textures such as $k\pi$ skyrmions, antiskyrmions or three-dimensional objects like chiral bobbers and hopfions can be regarded as an essential part of future research efforts.  

\section*{Data Availability}
Data sharing is not applicable to this article as no new data were created or analyzed in this study.

\begin{acknowledgments}
M.\ L.\ acknowledges the financial support by the German Science Foundation (Deutsche Forschungsgemeinschaft, DFG) through the research fellowship LO 2584/1-1. The preparation of this perspective article was partially supported by the NSF through the University of Illinois at Urbana-Champaign Materials Research Science and Engineering Center DMR-1720633 and was carried out in part in the Materials Research Laboratory Central Research Facilities, University of Illinois. 
\end{acknowledgments}

\bibliography{Skyrmion_literature}


\clearpage

\end{document}